\documentclass{PoS}
\usepackage{amssymb}   
\usepackage{amsmath}
\usepackage{subfigure}
\usepackage{bm}

\newcommand{\ajpk}{\ensuremath{A^{J/\psi K}}}
\newcommand{\ajppi}{\ensuremath{A^{J/\psi \pi}}}
\newcommand{\BplusDecay}{\ensuremath{B^+ \rightarrow J/\psi K^{+}}}
\newcommand{\BminusDecay}{\ensuremath{B^- \rightarrow J/\psi K^{-}}}
\newcommand{\BplusPiDecay}{\ensuremath{B^+ \rightarrow J/\psi \pi^{+}}}
\newcommand{\BminusPiDecay}{\ensuremath{B^- \rightarrow J/\psi \pi^{-}}}

\title{An improved measurement of direct $CP$ violation parameters in 
$B^\pm \rightarrow J/\psi K^{\pm}$ and $B^\pm \rightarrow J/\psi \pi^{\pm}$ decays}

\ShortTitle{Direct ${CP}$ violation in 
$B^\pm \rightarrow J/\psi K^{\pm}$ and $B^\pm \rightarrow J/\psi \pi^{\pm}$ decays.}

\author{\speaker{Iain Bertram}%
                \thanks{representing the D0 Collaboration.}\\
       Lancaster University\\
       E-mail: \email{i.bertram@lancaster.ac.uk}}

\abstract{We present a  measurement of the direct $CP$-violating charge asymmetry in $B^\pm$  mesons  
decaying  to $J/\psi K^{\pm}$ and $J/\psi \pi^{\pm}$ where $J/\psi$ decays to $\mu^+ \mu^-$, 
using 10.4 fb$^{-1}$ of proton-antiproton collisions collected by the D0 detector during Run II 
at the Fermilab Tevatron Collider. A difference in the yield of $B^-$ and  $B^+$ mesons in these decays 
is found by fitting to the difference between their reconstructed invariant mass distributions 
resulting in asymmetries of $\ajpk =\left[ \rm{0.59} \pm 0.36 \right]\%$, 
which is  the most precise measurement to date, and $\ajppi= \left[ \rm{-4.2} \pm  4.8 \right]\%$.
Both measurements are consistent with  standard model predictions. These 
measurements are combined with all previous measurements to form new world averages of \ajpk\ and \ajppi .
}

\FullConference{14th International Conference on B-Physics at Hadron Machines\\
		 April 8-12, 2013\\
		 Bologna, Italy}

\begin{document}

Currently all measurements of $CP$ violation, either in decay, mixing, or in the interference 
between the two, have been consistent with the presence of a single phase in the CKM matrix. 
The standard model predicts that for $b \rightarrow sc\bar{c}$ decays, the tree and penguin 
contributions have the same weak phase, and thus no direct $CP$ violation is expected in the 
decays of $B^\pm$ mesons to $J/\psi K^{\pm}$. 
Estimates of the effect of penguin loops~\cite{hou} show that there could be a small amount of 
direct $CP$ violation of up to ${\cal O}(0.3\%)$.
A measurement of a relatively large charge asymmetry would indicate the existence 
of physics beyond the standard model~\cite{hou, barger, wu}.
In the transition $b \rightarrow dc\bar{c}$, the tree and penguin contributions have 
different phases, and  there may be measurable levels of $CP$ violation in the 
decay $B^\pm \rightarrow J/\psi \pi^{\pm}$~\cite{bscc,hou2}.

The $CP$-violating charge asymmetry in the decays $B^\pm \rightarrow J/\psi K^{\pm}$ 
and $B^\pm \rightarrow J/\psi \pi^{\pm}$ are defined as 
\begin{align}
\ajpk  =  &\frac{\Gamma\left(\BminusDecay \right) - \Gamma\left(\BplusDecay \right)}{\Gamma\left(\BminusDecay \right) + \Gamma\left(\BplusDecay \right)},\\
\ajppi =  &\frac{\Gamma\left(\BminusPiDecay \right) - \Gamma\left(\BplusPiDecay \right)}{\Gamma\left(\BminusPiDecay \right) + \Gamma\left(\BplusPiDecay \right)}. 
\end{align}
Previous measurements of \ajpk \cite{belle2010,d02008,belle2008,babar2005,cleo2000} have been 
averaged by the Particle Data Group with the result 
$\ajpk = \left[  0.1 \pm 0.7 \right]\%$~\cite{pdg2012}. The most precise measurement of \ajpk\ 
was made by the Belle collaboration~\cite{belle2010}, with a total uncertainty of $0.54\%$. 
The most precise measurement of \ajppi\ was made by the LHCb collaboration~\cite{lhcb2012}, 
with a total uncertainty of $2.9\%$. The LHCb measurement is actually a measurement of the 
difference, $\ajppi - \ajpk$, and assumes that \ajpk\ is zero.

This Note presents a summary of the substantially improved measurements of \ajpk\ and \ajppi\ using 
the full Tevatron Run II data sample with an integrated luminosity of 10.4~fb$^{-1}$ which are described in detail in~\cite{d02013}. 

It is assumed that there is no production asymmetry between $B^+$ and $B^-$ mesons 
in proton-antiproton collisions. An advantage of these decay modes into $J/\psi X^\pm$ 
is that no assumptions on the $CP$ symmetry of subsequent charm decays need to 
be made.

These updated measurements of \ajpk\ and \ajppi\ make use of the methods for extracting 
asymmetries used in the analyses of the time-integrated flavor-specific semileptonic charge 
asymmetry in the decays of neutral $B$ mesons~\cite{d0assl,d0adsl}. We measure the 
raw asymmetries 
\begin{align}
A_{\rm raw}^{J/\psi K}  = &\frac{N_{J/\psi K^-} - N_{J/\psi K^+}  }{N_{J/\psi K^-} + N_{J/\psi K^+} },\\
A_{\rm raw}^{J/\psi \pi}  = &\frac{N_{J/\psi \pi^-} - N_{J/\psi \pi^+}  }{N_{J/\psi \pi^-} + N_{J/\psi \pi^+} },
\end{align}
where $N_{J/\psi K^-}$ ($N_{J/\psi K^+}$) is the number of reconstructed \BminusDecay\ 
(\BplusDecay ) decays, and $N_{J/\psi \pi^-}$ ($N_{J/\psi \pi^+}$) is the number of 
reconstructed \BminusPiDecay\ (\BplusPiDecay ) decays. The charge asymmetry in $B^\pm$ 
decays is then given by (neglecting any terms second-order or higher in the asymmetry) 
\begin{align}
\label{eq:asymm}
A^{J/\psi K}  = & A_{\rm raw}^{J/\psi K}  + A_{K},\\
A^{J/\psi \pi}  = & A_{\rm raw}^{J/\psi \pi} + A_\pi,
\end{align}
where $A_K$ is the dominant correction and is the reconstruction asymmetry between 
positively and negatively charged kaons in 
the detector. The correction $A_K$ is calculated using the measured kaon reconstruction asymmetry 
as described in~\cite{d0adsl}. As discussed later, data collected using regular 
reversals of magnet polarities results in no significant residual track reconstruction
asymmetries, and hence, no correction for tracking asymmetries or pion reconstruction 
asymmetries need to be applied, hence $A_\pi =0$.

The raw asymmetries are extracted by fitting the  data sample using an 
unbinned maximum likelihood fit. 

\begin{figure*}[ht]
        \centering
        \begin{minipage}[t]{.45\textwidth}
        \includegraphics[width=0.95\textwidth]{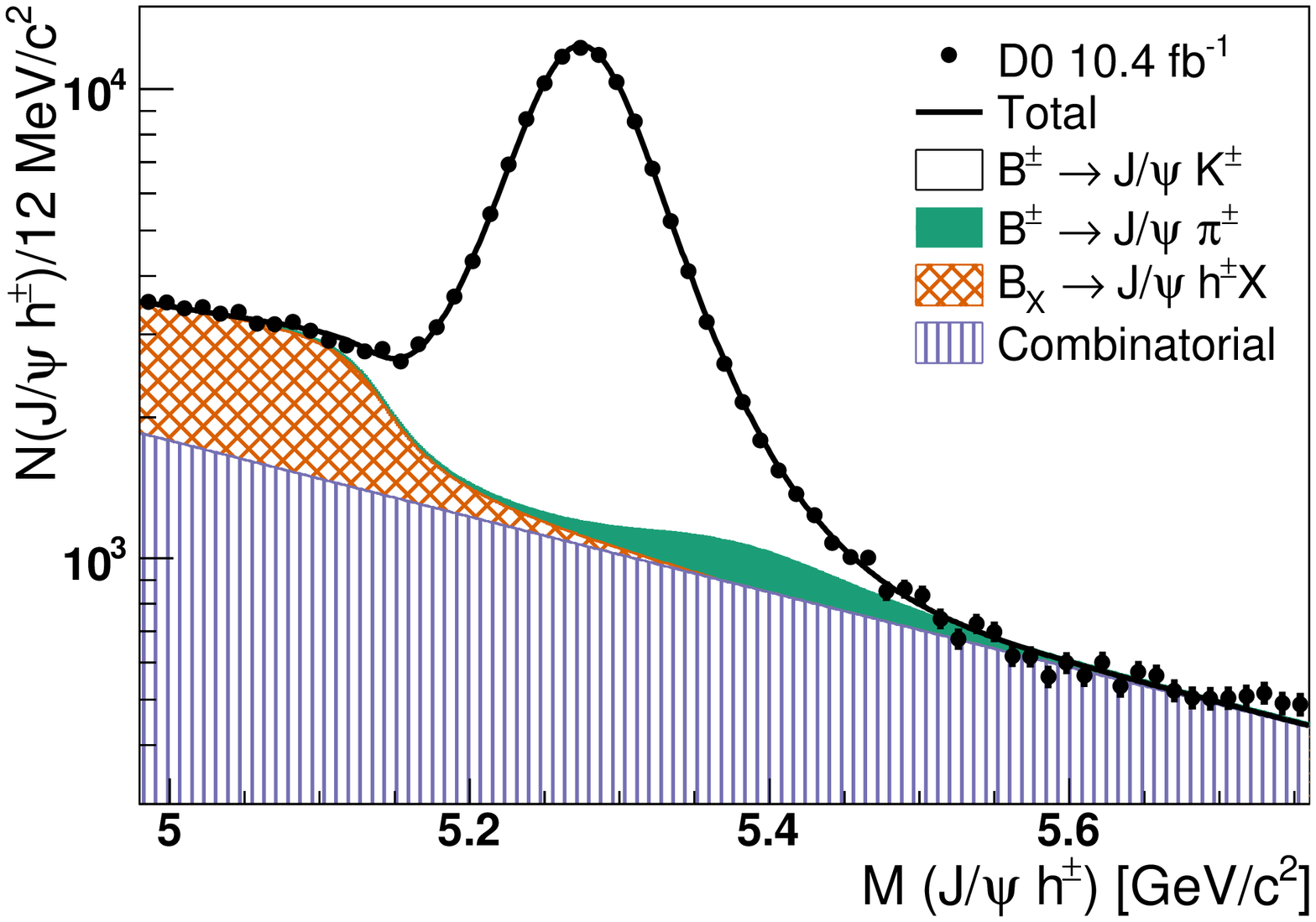}
        \centerline{Mass distribution}
        \end{minipage}
        \hspace{0.05\columnwidth}
        \begin{minipage}[t]{.45\textwidth}
        \includegraphics[width=0.95\textwidth]{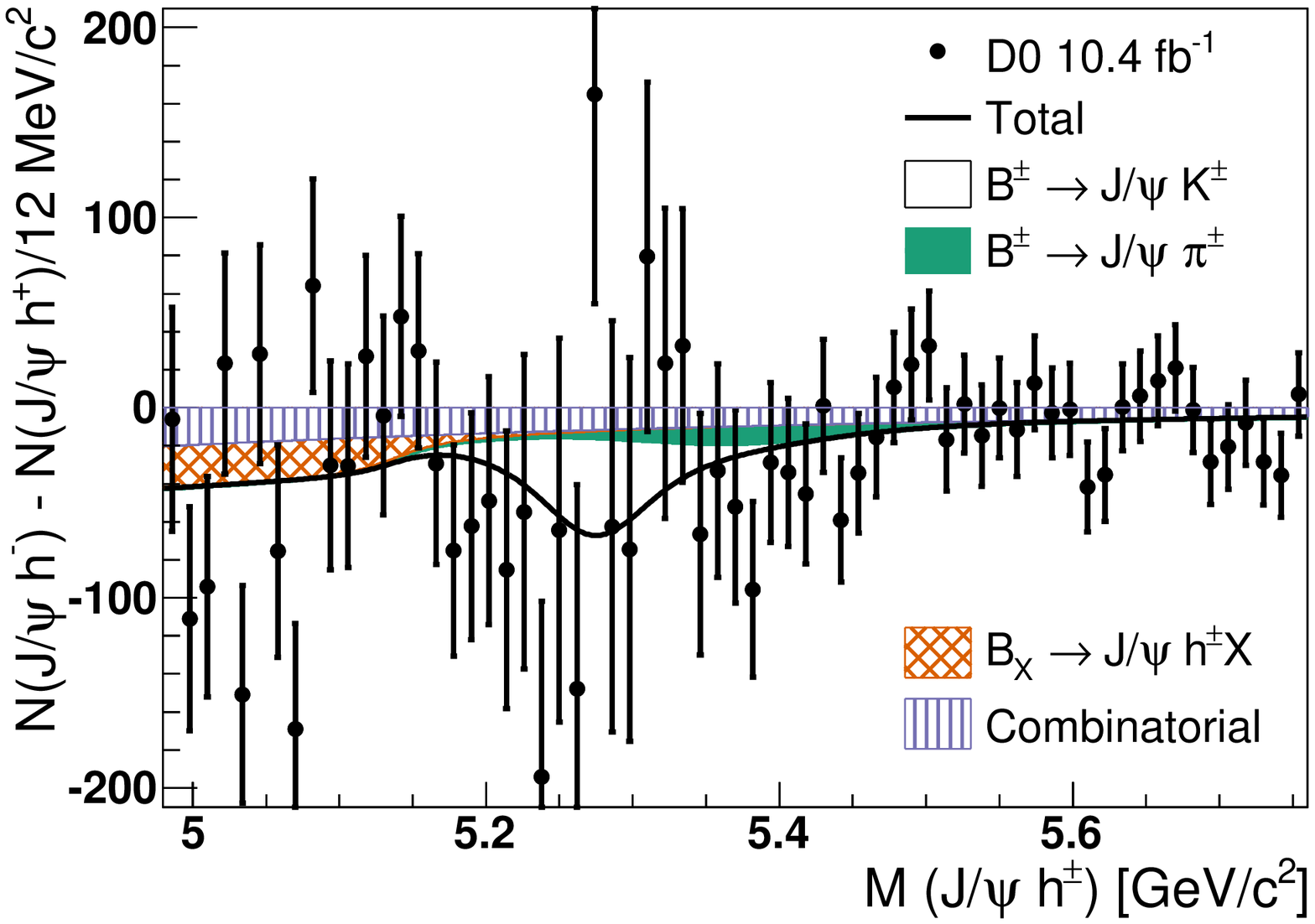}
        \centerline{Difference distribution}
        \end{minipage}
        \caption[]{The polarity-weighted $J/\psi h^\pm$ invariant mass distribution, where the $h^\pm$ is  
assigned the charged kaon mass. }
\label{fig:fit}
\end{figure*}

The number of signal candidates are extracted from the $J/\psi h^{\pm}$ mass 
distribution using an unbinned  maximum likelihood fit over a mass range 
of $4.98 <  M(J/\psi h^{\pm}) < 5.76$~GeV$/c^2$.
The dominant peak consists of the overlap of the $B^\pm \rightarrow J/\psi K^\pm$ 
and the $B^\pm \rightarrow J/\psi \pi^\pm$ (where the $\pi^\pm$ is mis-identified 
as a $K^\pm$) components. The mis-identified $B^\pm \rightarrow J/\psi \pi^\pm$ 
decay mode appears as a small peak shifted to a slightly higher mass than the $B^\pm$.
The $B^\pm \rightarrow J/\psi K^\pm$ signal peak is modeled by two Gaussian functions 
constrained to have the same mean but, with different widths and normalizations to 
model the detector's mass resolution. Taking account the D0 momentum scale, 
the mean is found to be consistent with the PDG average of the $B^\pm$ meson mass.
To obtain a good fit to the data, the widths  
have a linear dependence on the kaon energy. 
We assume that the mass distribution of the $B^\pm \rightarrow J/\psi \pi^\pm$ 
is identical to that of $B^\pm \rightarrow J/\psi K^\pm$, if the correct hadron 
mass is assigned. To  model the $J/\psi \pi^\pm$  mass distribution, $G_\pi(m)$, 
the $J/\psi \pi^\pm$ signal peak is transformed by assigning the pion track the 
charged kaon mass.
The resulting $J/\psi h^\pm$ polarity-weighted 
invariant mass distribution is shown in Fig.~\ref{fig:fit} (where $h^{\pm}$ is any charged hadron).  
The $B^\pm \rightarrow J/\psi K^\pm$ signal contains $105562 \pm 370 \thinspace (\mbox{stat})$ events,  
and the $B^\pm \rightarrow J/\psi \pi^\pm$ signal contains $3110 \pm 174 \thinspace (\mbox{stat})$ events.

The invariant mass distribution of the differences, $N(J/\psi h^-) - N(J/\psi h^+)$, 
are also  shown in Fig.~\ref{fig:fit} with a resulting $\chi^2$ of 58.5 for 61 degrees of 
freedom. The resulting raw asymmetries are extracted from the data are (including the effect of
systematic uncertainties on the fitting procedure):
\begin{align}
   A_{\rm raw}^{J/\psi K} = {}&  \left[ -0.46 \pm 0.36 \thinspace (\mbox{stat}) \pm 0.046 \thinspace (\mbox{syst}) \right]\%, \\
   A_{\rm raw}^{J/\psi \pi} = {}&  \left[ -4.2 \pm 4.4 \thinspace (\mbox{stat}) \pm 1.82 \thinspace (\mbox{syst}) \right]\%.
\end{align}

The  raw asymmetry for \ajpk\ is corrected by 
\begin{equation}
A_K = \left[ 1.046 \pm 0.043 \thinspace (\mbox{syst})  \right] \%.
\end{equation}
Resulting in final asymmetries of
\begin{align}
\ajpk =& \left[ \rm{0.59} \pm 0.36 \thinspace(\text{stat}) \pm 0.08 \thinspace(\text{syst}) \right]\%, \label{ajkresult}\\
\ajppi=& \left[ \rm{-4.2} \pm 4.4 \thinspace(\text{stat}) \pm 1.8 \thinspace(\text{syst}) \right]\%. \label{ajpiresult}
\end{align}
This is the most precise measurement of \ajpk\ to date and is a reduction in 
uncertainty by approximately a factor of two from the previous D0 result~\cite{d02008}.

\begin{figure*}[ht]
        \centering
        \begin{minipage}[t]{.45\textwidth}
        \includegraphics[width=0.95\textwidth]{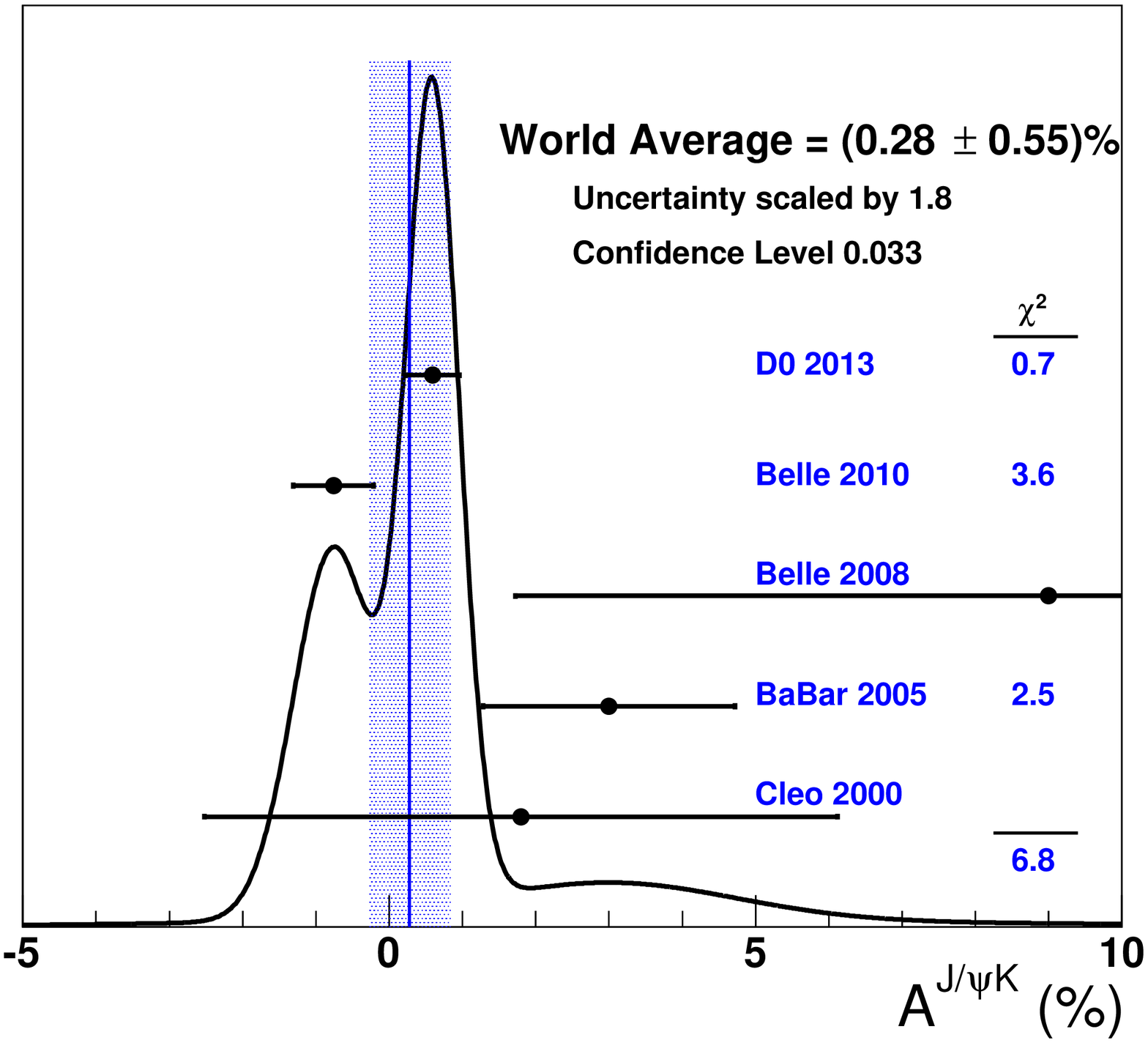}
        \centerline{\ajpk\ combination}
        \end{minipage}
        \hspace{0.05\columnwidth}
        \begin{minipage}[t]{.45\textwidth}
        \includegraphics[width=0.95\textwidth]{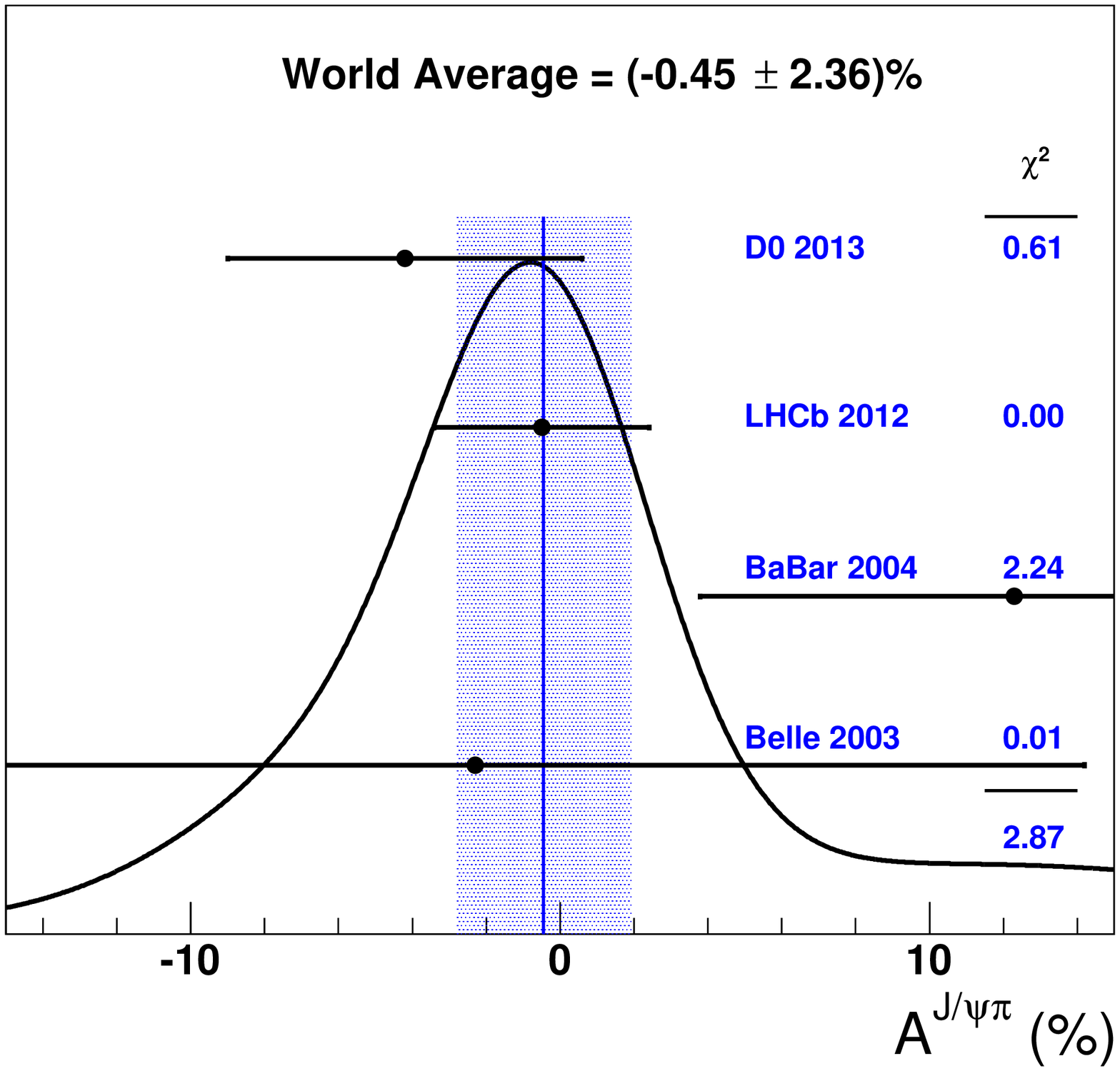}
        \centerline{\ajppi\ combination}
        \end{minipage}
        \caption[]{Combination of all measurements of \ajpk\ and \ajppi\ made using the method used by the PDG~\cite{pdg} (see text). }
\label{fig:combination}
\end{figure*}

The D0 measurements of \ajpk\ and \ajppi\ can be combined with all other measurements to form updated world averages 
(Fig.~\ref{fig:combination}). I use a simple weighted average, assuming that the  measurements are fully independent. For \ajpk\ results from Belle~\cite{belle2010,belle2008}, BaBar~\cite{babar2005} and Cleo~\cite{cleo2000} are combined with the D0 result. The resulting $\chi^2$ for the three most precise measurements is 6.8, indicating that the measurements are not very consistent. The resulting error is then scaled by the square root of the $\chi^2$ per degree of freedom, 1.8, giving 
\begin{equation}
\ajpk(\rm{WA}) = (0.28 \pm 0.55) \%.
\end{equation}
For \ajppi\ results from LHCb~\cite{lhcb2012}, BaBar~\cite{babar2004} and Belle~\cite{belle2003} are combined with the D0 result resulting in 
\begin{equation}
\ajpk(\rm{WA}) = (-0.45 \pm 2.36) \%.
\end{equation}
Both results are consistent with the standard model predictions.

\end{document}